\documentclass[conference]{IEEEtran}
\IEEEoverridecommandlockouts

\usepackage{ifpdf}
\usepackage{cite}
\usepackage{placeins}
\usepackage[pdftex]{graphicx}
\usepackage{array}
\usepackage{caption}
\usepackage{subcaption}
\usepackage{fixltx2e}
\usepackage{stfloats}
\usepackage{amsthm}
\usepackage{amsmath}
\usepackage{bbm}
\usepackage{amsfonts}
\usepackage{amssymb}
\usepackage[T1]{fontenc} 
\usepackage[cmintegrals]{newtxmath}
\usepackage{bm} 
\usepackage[all]{xy}
\usepackage{enumitem}
\usepackage{float}
\usepackage[usenames, dvipsnames]{color}
\usepackage{ifpdf}
\usepackage{cite}
\usepackage{array}
\usepackage{mathtools}

\makeatletter
\def\widebreve{\mathpalette\wide@breve}
\def\wide@breve#1#2{\sbox\z@{$#1#2$}%
	\mathop{\vbox{\m@th\ialign{##\crcr
				\kern0.08em\brevefill#1{0.8\wd\z@}\crcr\noalign{\nointerlineskip}%
				$\hss#1#2\hss$\crcr}}}\limits}
\def\brevefill#1#2{$\m@th\sbox\tw@{$#1($}%
	\hss\resizebox{#2}{\wd\tw@}{\rotatebox[origin=c]{90}{\upshape(}}\hss$}
\makeatletter

\usepackage{url}
\usepackage{mathtools}
\usepackage[dvipsnames]{xcolor}
\usepackage{multirow}
\usepackage{mwe} 
\usepackage{times}
\usepackage{wasysym}
\usepackage{epsfig}
\usepackage{multirow}
\usepackage{tabularx}
\usepackage{graphicx}
\usepackage{color}
\usepackage{xspace}
\usepackage{thumbpdf}
\usepackage{listings}
\usepackage{verbatim}
\usepackage{outlines}
\usepackage{textgreek}
\usepackage{booktabs}
\usepackage{amsfonts}
\usepackage{colortbl}
\usepackage{multicol}
\usepackage{multirow}
\usepackage{balance}
\usepackage{makecell}
\usepackage{array}
\usepackage[normalem]{ulem}
\usepackage{mathtools,amssymb}
\usepackage{float} 
\usepackage{booktabs} 
\usepackage{wrapfig}
\usepackage{subfloat}

\usepackage{diagbox}
\usepackage{cleveref}
\usepackage[]{algorithm2e}

\theoremstyle{definition}

\usepackage[noend]{algpseudocode}

\usepackage{diagbox}

\theoremstyle{definition}

\begin{document}
\title{Optimal Routing and Link Configuration for Covert Heterogeneous Wireless Networks\\}

 \author{%
   \IEEEauthorblockN{Amna Gillani\IEEEauthorrefmark{1},
                     Beatriz Lorenzo\IEEEauthorrefmark{1},
                     Majid Ghaderi\IEEEauthorrefmark{2},
                     Fikadu Dagefu\IEEEauthorrefmark{3}, and
                     Dennis~Goeckel\IEEEauthorrefmark{1}} \\
   \IEEEauthorblockA{\IEEEauthorrefmark{1}%
                     Department of Electrical and Computer Engineering,
                     University of Massachusetts Amherst,
                     Amherst, MA, 01003 USA\\
                     \{agillani, blorenzo, dgoeckel\}@umass.edu}
    \IEEEauthorblockA{\IEEEauthorrefmark{2}%
                     Department of Computer Science, University of Calgary, Calgary, AB T2N 1N4, Canada \\
                     mghaderi@ucalgary.ca}
   \IEEEauthorblockA{\IEEEauthorrefmark{3}%
                     U.S. Army Combat Capabilities Development Command (DEVCOM), Army Research Laboratory, Adelphi, MD 20783 USA \\
                     fikadu.t.dagefu.civ@army.mil}
    \thanks{This work has been submitted to the IEEE for possible publication. Copyright may be transferred without notice, after which this version may no longer be accessible. 
    This work was supported in part by the National Science Foundation under ECCS-2148159.  Research was also sponsored in part by the Army Research Laboratory and was accomplished under Cooperative Agreement Number W911NF-23-2-0014.
The views and conclusions contained in this document are those of the
authors and should not be interpreted as representing the official policies,
either expressed or implied, of the Army Research Laboratory or the U.S.
Government. The U.S. Government is authorized to reproduce and distribute
reprints for Government purposes notwithstanding any copyright notation
herein.}
 }
	
\maketitle

\begin{abstract}
Nodes in contemporary radio networks often have multiple interfaces available for communication: WiFi, cellular, LoRa, Zigbee, etc.  This motivates understanding both link and network configuration when multiple communication modalities with vastly different capabilities are available to each node.  In conjunction, covertness or the hiding of radio communications is often a significant concern in both commercial and military wireless networks.  We consider the optimal routing problem in wireless networks when nodes have multiple interfaces available and intend to hide the presence of the transmission from attentive and capable adversaries.  We first consider the maximization of the route capacity given an end-to-end covertness constraint against a single adversary, and we find a polynomial-time algorithm for optimal route selection and link configuration.  We further provide optimal polynomial-time algorithms for two important extensions:  (i) statistical uncertainty during optimization about the channel state information for channels from system nodes to the adversary; and, (ii) maintaining covertness against multiple adversaries.  Numerical results are included to demonstrate the gains of employing heterogeneous radio resources and to compare the performance of the proposed approach versus alternatives.
\end{abstract}
\begin{IEEEkeywords}
heterogenerous networks, covert communications
\end{IEEEkeywords}
\section{Introduction}
\label{intro}

Nodes in wireless networks are often equipped with different sets of radios operating at different parts of the electromagnetic spectrum \cite{fikadu_tifs_2023}.  These communication modalities, referring to various wireless communication technologies, offer distinct advantages depending on the situation. Some radios provide long-range and low data rate performance while others might provide high data rates in short-range line of sight communication \cite{kong_wcnc2022}. Network performance can be significantly enhanced by optimally leveraging available modalities at each node in multi-hop routes, considering local wireless conditions, interference from other routes, and the presence of potential adversaries.

Security and privacy are critical aspects of contemporary commercial and military communication systems.  Security typically focuses on protecting the content of a message through encryption or information-theoretic (e.g., wiretap \cite{wyner_wiretap}) approaches. But the stronger security of hiding the {\em presence} of the message is often desirable for different use cases.  An example medical application is where the signal from an implanted medical device must be hidden to protect the patient's privacy.  Another relevant application is in military communications where the presence of radio signals could reveal network activity making the network vulnerable to adversarial attacks.  The fundamental limits of communication while keeping the signal hidden from an attentive and capable adversary were introduced in \cite{bash2013limits} and are now known as ``covert'' communications in the literature. We present a detailed literature review in Section \ref{sec:related}. 

We consider the route optimization problem in a heterogeneous network with nodes having multiple radio interfaces where the optimal combination of modalities for each hop in the route is selected, with the goal of maximizing the end-to-end capacity under an end-to-end covert communication constraint.  An analogous optimization for homogeneous radios was performed in \cite{covert_routing_2018} where efficient polynomial-time algorithms were obtained that either maximize capacity or minimize the latency of communication from a source to a destination.  However, the problem here is more challenging since the multiple interfaces preclude the approach employed in \cite{covert_routing_2018}.  Thus a novel approach, based on first principles for simultaneous optimal modality selection for each hop and efficient route selection, is required.

Optimal multi-hop routing in networks of nodes with multiple communication modalities available was first considered in \cite{fikadu_tifs_2023} where a covertness constraint is also enforced and extended to decentralized routing in \cite{fikadu_wcl, fikadu_wisec_2024}.  However, in \cite{fikadu_tifs_2023,fikadu_wcl, fikadu_wisec_2024}, a single mode is selected for each hop, and, importantly, a fixed per-hop covertness constraint rather than end-to-end covertness constraint is enforced, hence simplifying the optimization problem in \cite{fikadu_tifs_2023} by allowing for link costs to be readily assigned in each of the three scenarios considered and shortest path or widest path routing is readily employed.  The derivation of a polynomial-time algorithm in Section \ref{sec:optimization} will prove to be more subtle and non-trivial.

We make the following contributions:  
\begin{itemize} 
\item{Develop an optimal polynomial-time routing and link optimization algorithm for wireless networks with multimodal nodes, maximizing end-to-end capacity while maintaining covertness against an attentive adversary monitoring all links.} 
\item{Extend the model to address:}
\begin{itemize} 
\item the case where multiple observers optimally combine their observations to attempt network detection.
\item statistical uncertainty in channel gains observed by adversaries. 
\end{itemize}
\end{itemize}
Section \ref{sec:related} reviews related work.  Section \ref{sec:model} presents the system model and metrics.  Section \ref{sec:optimization} presents the optimization in the case of a single adversary with known channel gains (i.e., known channel state information (CSI)) from system nodes to this adversary.  Section \ref{sec:extensions} presents the extensions to multiple adversaries and unknown CSI, while Section \ref{sec:numerical} presents numerical results.  Section \ref{sec:conclusion} presents the conclusions.
\section{Related Work}
\label{sec:related}
Fueled by exponential growth in technology, today's world has become progressively interconnected and diverse. By 2025, the International Data Corporation estimates there will be 55.7 billion connected Internet of Things (IoT) devices generating nearly 80 zettabytes of data \cite{hojlo2021future}. With this growing demand, the network must provide end-to-end optimal bandwidth, low latency, and secure communication. Traditional networks optimized for homogeneous traffic face unprecedented challenges to meet these demands cost-effectively. As a solution, heterogeneous networks (HetNets) are deployed to improve capacity and enhance network coverage. A HetNet comprises nodes with diverse wireless technologies such as LTE, 5G, and 3G, allowing devices to switch between networks for seamless connectivity, enhanced performance, and improved reliability.

The problem of resource allocation and mode selection in HetNets, aimed at optimizing communication mode choices and balancing resource utilization, has been studied extensively in \cite{feng2020smart, algedir2020energy, sultan2015mode, xu2021survey, amiri2018machine, hu2011hetnets, mehrpouyan2015hybrid}. While HetNets improve performance and reliability, the growing complexity introduces new vulnerabilities, particularly at the physical layer. However, these studies often overlook the physical layer security of the network. Modern communication networks prioritize security, which typically involves ensuring that messages are indecipherable by adversaries, known as the wiretap problem. The heterogeneous wiretap problem was first studied by Lv et al. in \cite{lv2015secrecy} for a two-tier downlink heterogeneous network with a single adversary present in each cell, followed by the study of Wu et al. in \cite{wu2015secrecy, wu2018survey} on secrecy outage probability for k-tier homogeneous and heterogeneous networks. A dynamic coordinated scheme to enhance secure communication coverage was proposed by Xu et al. in \cite{xu2016enhancing}. The wiretap problem for physical-layer security has been further explored in \cite{wang2016physical}, \cite{wang2018secure}, \cite{liu2016physical} and \cite{xu2018security}.

However, even the detection of the presence of a message can compromise security and privacy. For instance, in military scenarios, the consequences of the potential detection of radio signals make the concealment of radio networks crucial. This requirement for security is addressed through covert communication, which involves sending messages in a manner that hides even the existence of the message from unintended observers.

Bash et al. defined the fundamental limits of covert communication for additive white Gaussian noise (AWGN) channels in \cite{bash2012square, bash2013limits}, introducing the square-root law (SRL), subsequently extended to other channel models and key size requirements in \cite{jaggi_isit, ligong_isit, ligong_covert2016, bloch_isit, bloch15covert}. Developing from this foundation, significant advances in covert communication were made to provide a positive rate by addressing the unknown channel state information of adversaries, as demonstrated by Lee et al. in \cite{lee2014achieving, lee2015achieving} and by introducing an uninformed jammer (as shown by Sobers et al. in \cite{sobers2017covert}).  Extensions to the multiple-access network have also been considered (e.g., \cite{arumugam2019}).  Comprehensive reviews at various stages of the development of the covert communications literature can be found in \cite{bash2015,yan2019,chen2023}.

While substantial progress has been made in covert communication research, much of it has focused on single-link optimization rather than network-level approaches. Dehghan et al. in \cite{dehghan2011minimum} and Ghaderi et al. in \cite{ghaderi2014minimum} have studied a network-level approach for minimum energy routing problem with channel variation and in the presence of a friendly jammer, respectively. Sheikholeslami et al. in \cite{covert_routing_2018} further studied routing in covert wireless networks with multiple collaborating adversaries, where minimum delay and maximum throughput covert routing schemes were introduced. However, these approaches only consider homogeneous networks. Other works on homogeneous covert networks include \cite{wang2019secrecy}, \cite{ma2021covert}, \cite{zheng2021wireless}, and \cite{im2020mobility}.  Most previous works in covert communication assumed a single-link or homogeneous network.

In \cite{yang2021mode}, Yang et al. studied mode selection and cooperative jamming for covert communication. In \cite{kong2024covert, kong_wcnc2022} and \cite{fikadu_tifs_2023}, Kong et al. and Aggarwal et. al. in \cite{aggarwal2024covert} considered in detail methods for optimizing performance (capacity, latency) under a covertness constraint for multiple-modality nodes. In \cite{fikadu_wcl}, Kong et al. and \cite{fikadu_wisec_2024}, Kim et al. introduced a reinforcement learning-based approach to finding the optimal route in a heterogeneous network. However, these works primarily focus on mode selection and enforce a fixed covertness constraint per link by equally dividing the end-to-end covertness constraint across the path. We employ modes simultaneously and enforce an end-to-end covertness constraint, optimally allocating it across the links of the chosen path. Due to these differences, we observe that the approach in \cite{kong2024covert} that allowed link costs to be readily determined to support efficient end-to-end routing does not apply; rather, we need to redesign the network optimization from the first principles.
\section{System Model, Definitions, and Metrics}
\label{sec:model}
\subsection{Network and Link Model}

Consider a network of \(N\) friendly nodes, ${\cal T}=\{T_1, T_2, \ldots, T_N\}$, where each system node has $M \geq 1$ modalities for transmission/reception and each adversary (often termed ``Willie'' in the covert communications literature) has a receiver for each of the $M$ modes\footnote{Optimization in the case where each node has a subset of the modalities available follows similarly by setting the channel gain for a given mode to zero if that mode is not available for direct communication between a given pair of nodes.}.  Also present is a set of adversaries ${\cal W}=\{W_1, W_2, \ldots, W_K\}$, whose goal is to detect any communication in the network.  Between each pair of nodes and on each mode exists a multipath fading channel subject to a fading gain, which may or may not be known when doing route planning, pathloss with exponent $\alpha$, and additive white Gaussian noise (AWGN).  

The transmission is done using standard Gaussian codebooks \cite{bash2013limits}. 
On mode $m$, $m=1,2,\ldots, M$, a transmitter $X$ will employ a message $[f_{m,1}, f_{m,2}, \ldots, f_{m,n}]$ drawn from a unit-energy Gaussian codebook (i.e., with symbols $\sim {\cal N}(0,1)$) based on the message, where $n$ is the length of a codeword.  The signal that (friendly) receiver $Y$ receives on  mode $m$ is given by:
\begin{equation}
    Z_{m,j}^{(Y)} = \frac{g_{(X,Y),m} \sqrt{P_{X,m}}}{d_{X,Y}^{\alpha/2}} f_{m,j} + N^{(Y)}_{m,j}, ~~~j=1,2,\ldots n
\end{equation}
where $g_{(X,Y),m}$ is the fading gain between $X$ and $Y$ on mode $m$, which is generally assumed to be known since it is between friendly nodes and can be obtained via pilot signaling, $P_{X,m}$ is the transmit power of node $X$ on mode $m$, $d_{X,Y}$ is the distance between node $X$ and $Y$, and $N^{(Y)}_{m,j}\sim {\cal N}(0,\sigma_{Y,m}^2)$ is the AWGN at the receiver for mode $m$ at node $Y$.  The signal that Willie $W_k$,$k=1,2,\ldots K$, receives on mode $m$ is given by:
\begin{equation}
    Z_{m,j}^{(W_k)} = \frac{g_{(X,W_k),m} \sqrt{P_{X,m}}}{d_{X,W_k}^{\alpha/2}} f_{m,j} + N^{(W_k)}_{m,j}, ~~~j=1,2,\ldots n
\end{equation}
where $g_{(X,W_k),m}$ and $d_{X,W_k}$ are defined analogously to above, and $N^{(W_k)}_{m,j} \sim {\cal N}(0,\sigma_{W_k,m}^2)$.

\subsection{Adversary Model}

The adversaries ${\cal W}=\{W_1, W_2, \ldots, W_K\}$ will attempt to detect whether there is a communication in the network from the source to the destination.  To model a strong adversary, we assume the following: (1) adversaries know the parameters of potential transmission, including the route from the source to the destination and the power levels used by each node on each modality; (2) adversaries are aware of the transmission schedule, allowing them to detect when transmissions occur on specific modes employed by a given node; (3) multiple adversaries collaborate and optimize their decision-making process based on joint observations.

As in a regular transmission system, the codebook is shared among the friendly nodes but unknown to the adversaries \cite{bash2013limits}.  Hence the codebook forms the shared secret that provides the advantage receivers need over adversaries for covert communications in the pessimistic (and likely) case when an advantage in the wireless environment cannot be assumed a priori.

\subsection{Covertness Constraint}

We adopt the Kullback-Leibler divergence \cite{cover1999elements} as the covertness criterion \cite{bloch15covert}.  That is, covertness at level $\epsilon$ is achieved on the route from the source to the destination if 
\begin{equation}
    {\cal D}({\cal Q}_0 || {\cal Q}_1) \leq \epsilon
\end{equation}
where ${\cal D}(p || q)$ is the Kullback-Leibler divergence between probability distributions $p$ and $q$, ${\cal Q}_0$ is the distribution of the (vector) observations across all adversaries on all modes and all hops for all \(n\) transmissions when there is not a message moving across the network, and ${\cal Q}_1$ is the analogous (vector) distribution when there is a message in the network.

The Kullback-Leibler divergence is a standard measure of the difference between two probability distributions.  Due to Pinsker's inequality \cite{cover1999elements}, it has operational significance in the covert communications scenario through the lower bound $P_E \geq \frac{1}{2} - \frac{1}{2} \sqrt{\frac{1}{2} {\cal D}({\cal Q}_0 || {\cal Q}_1)} $, where $P_E$ is the probability of error in the decision at the adversaries when a transmission in the network happens with probability $\frac{1}{2}$.  That is, if ${\cal D}({\cal Q}_0 || {\cal Q}_1) \leq \epsilon$, then the adversaries are restricted to $P_E \geq \frac{1}{2} - \frac{\epsilon}{2 \sqrt{2}}$.

The constraint on adversarial detection of a codeword traversing the network can be directly translated into a constraint for each code symbol, as follows.
From the codebook construction and the network model, observations of the adversaries when conditioned on the channel gains and distances will be independent and identically distributed across the sequence of symbol slots $j=1,2,\ldots,n$ corresponding to a codeword.  Hence, define $\mathbb{P}_0$ as the distribution of the (vector of) observations for a given codeword position across adversaries, links, and modes in the network when there is no transmission traversing the network, and $\mathbb{P}_1$ as the analogous distribution when there is a transmission present.
Then, using the chain rule \cite{cover1999elements},
\[\mathcal{D}({\cal Q}_0 || {\cal Q}_1) = \mathcal{D}(\mathbb{P}_0^n||\mathbb{P}_1^n)=n\mathcal{D}(\mathbb{P}_0 || \mathbb{P}_1).\]
The covertness criteria is given by:
\[\mathcal{D}(\mathbb{P}_0 || \mathbb{P}_1) \le \delta = \frac{\epsilon}{n}\]
where $\delta$ is the covertness criterion per symbol, which will be employed extensively in successive sections.

\subsection{Optimization Criterion and Network State Knowledge}

We will maximize the capacity achieved on the route from the source to the destination while maintaining the covertness constraint.  Since the route capacity is limited by the minimum capacity of any link in the route, we will seek a route that has the maximum value for the minimum capacity of any link along the route.

The optimization assumes known locations of system nodes and adversaries, a scenario plausible when avoiding detection by a stationary radio tower. When adversary locations are unknown, an approach similar to the secrecy approach taken in \cite{ghaderi2014minimum} that defends against a set of adversary locations can exploit the multiple-adversary work of Section \ref{sec:extensions}; the numerical results of Section \ref{sec:numerical} indicate the impact on the network throughput.  The initial algorithm derived in Section \ref{sec:optimization} will assume that all channel gains are known when performing the optimization, but in Section \ref{sec:extensions}, we will show how the assumption of knowledge of the channel gain to the adversary is readily relaxed under statistical uncertainty (e.g., Rayleigh fading) due to the covertness bound only depending on the second moment of the received power.
\section{Optimization} \label{sec:optimization}

Per Section \ref{sec:model}, we seek to find the path from a source node $S$ to a destination node $D$ that maximizes the route capacity while meeting the end-to-end covertness constraint.  Let a path from $S$ to $D$ be denoted by $\Pi=(l_1, l_2, \ldots, l_H)$, where $H$ is the number of hops on the path, and $l_i=(S_i,D_i)$ is the $i^{th}$ link along the path.

To address the network optimization problem, we first consider the maximization of the capacity of a single multi-mode link in the presence of a single adversary Willie - a result of independent interest.  This yields the optimal combination of modes and resulting capacity for the link as a function of the covertness of the link.  This is then employed to find an optimal (maximum capacity) path in the network, which we show can be done in polynomial time via standard shortest-path algorithms.  

\subsection{Optimization of the Sum Capacity for a Single Multi-Mode Link in the Presence of a Single Willie}
\label{sec:single_Willie}

Consider the optimization of mode usage for link $i$ from $S_i$ to $D_i$.  In this section, we will maximize the sum capacity over all modes subject to a covertness criterion $\delta_i$ at a single Willie $W_k$.  That is, with $C_m$ the capacity of mode $m$, the problem is:
\begin{equation}
\mbox{max}_{P_{S_i,1}, P_{S_i,2},\ldots, P_{S_i,M}} \sum_{m=1}^M C_m
\end{equation}
such that
\begin{equation*}
{\cal D}({\mathbb P}^{(i)}_0 || {\mathbb P}^{(i)}_1) \leq \delta_i
\end{equation*}
where: (i) ${\mathbb P}^{(i)}_0$ and ${\mathbb P}^{(i)}_1$ are the distributions of the (vector) observations at the adversary from the modes of link $i$; (ii) $\delta_i$ is the per-symbol covertness constraint on link $i$, which is a network optimization parameter that will be found later.

Incorporating the details of the model given in Section \ref{sec:model} yields:
\begin{equation*}
\displaystyle \mbox{max}_{P_{S_i,1}, P_{S_i,2},\ldots, P_{S_i,M}} \frac{1}{2}\sum_{m=1}^M \log \left (1 + \frac{g^2_{(S_i,D_i),m} P_{S_i,m}}{\sigma^2_{D_i,m} d_{S_i,D_i}^{\alpha}} \right )
\end{equation*}
and, exploiting the well-known fact that the Kullback-Leibler divergence is locally quadratic in the received power \cite{kullback1959}, it can be shown by generalizing the results of \cite{bash2013limits,covert_routing_2018} to multiple modes that
\begin{equation*}
\sum_{m=1}^M \left (\frac{g^2_{(S_i,W_k),m}P_{S_i,m}}{\sigma^2_{W_k,m} d_{S_i,W_k}^{\alpha}} \right)^2 \leq \delta_i
\end{equation*}
is sufficient to maintain covertness at level $\delta_i$.

Since the power will be small for covert communications at practical blocklengths $n$, we can linearize the log in the capacity formula \cite{covert_routing_2018}, which yields the optimization problem:
\begin{equation}
\mbox{max}_{P_{S_i,1}, P_{S_i,2},\ldots, P_{S_i,M}} \sum_{m=1}^M  \frac{g^2_{(S_i,D_i),m} P_{S_i,m}}{2\sigma^2_{D_i,m} d_{S_i,D_i}^{\alpha}} 
\label{eq:linearized_capacity}
\end{equation}
such that
\begin{equation*}
\sum_{m=1}^M \left (\frac{g^2_{(S_i,W_k),m}P_{S_i,m}}{\sigma^2_{W_k,m} d_{S_i,W_k}^{\alpha}} \right)^2 \leq \delta_i.
\end{equation*}
Since the maximum capacity will be obtained for equality in the constraint equation, this problem is amenable to solution by Lagrange multipliers.  Define the Lagrangian:
\begin{eqnarray*}
    {\cal L}((P_{S_i,1}, P_{S_i,2},\ldots, P_{S_i,M}),\lambda) = ~~~~~~~~~~~~~~~~~~~~~ \\ \sum_{m=1}^M  \frac{g^2_{(S_i,D_i),m} P_{S_i,m}}{2\sigma^2_{D_i,m} d_{S_i,D_i}^{\alpha}} + \lambda \left (\delta_i - \sum_{m=1}^M \left (\frac{g^2_{(S_i,W_k),m} P_{S_i,m}}{\sigma^2_{W_k,m} d_{S_i,W_k}^{\alpha}} \right)^2 \right ).
\end{eqnarray*}
Taking the partials with respect to $P_{S_i,1}, P_{S_i,2}, \ldots P_{S_i,M}$ yields:
\begin{eqnarray*}
\frac{\partial}{\partial P_{S_i,j}}  {\cal L}((P_{S_i,1}, P_{S_i,2},\ldots, P_{S_i,M}),\lambda) = ~~~~~~~~~~~~~~~~~~~~ \\ 
\frac{g^2_{(S_i,D_i),j}}{2\sigma^2_{D_i,j} d_{S_i,D_i}^{\alpha}} - 2 \lambda \frac{g^4_{(S_i,W_k),j} P_{S_i,j}}{\sigma^4_{W_k,j} d_{S_i,W_k}^{2\alpha}}, ~~~~j=1,2,\ldots,M.
\end{eqnarray*}
Setting the partials equal to zero to solve for the necessary conditions on $P_{S_i,1}, P_{S_i,2}, \ldots P_{S_i,M}$ results in:
\begin{equation*}
P_{S_i,j} = \frac{1}{4 \lambda} \frac{d_{S_i,W_k}^{2 \alpha}}{d_{S_i,D_i}^{\alpha}} \frac{g^2_{(S_i,D_i),j} \sigma^4_{W_k,j}}{g^4_{(S_i,W_k),j}\sigma^2_{D_i,j}}.
\end{equation*}
A quick check shows that the power on a mode is proportional to the correct quantities.  The power is larger for noisier modes at Willie, less noisy modes at the destination, modes with large gains to the destination, and modes with small gains to Willie.  Substituting into the constraint equation yields:
\begin{eqnarray*}
    \sum_{m=1}^M \frac{g^4_{(S_i,W_k),m}}{\sigma^4_{W_k,m} d_{S_i,W_k}^{2\alpha}}P^2_{S_i,m} = \delta_i \\
    \sum_{m=1}^M \frac{g^4_{(S_i,W_k),m}}{\sigma^4_{W_k,m} d_{S_i,W_k}^{2\alpha}} \left (\frac{1}{4 \lambda} \frac{d_{S_i,W_k}^{2 \alpha}}{d_{S_i,D_i}^{\alpha}} \frac{g^2_{(S_i,D_i),m} \sigma^4_{W_k,m}}{g^4_{(S_i,W_k),m}\sigma^2_{D_i,m}} \right)^2 = \delta_i \\
    \sum_{m=1}^M \frac{1}{16 \lambda^2} \frac{d_{S_i,W_k}^{2 \alpha}}{d_{S_i,D_i}^{2 \alpha}} \frac{g^4_{(S_i,D_i),m} }{g^4_{(S_i,W_k),m}} \frac{\sigma^4_{W_k,m}}{\sigma^4_{D_i,m}}
    = \delta_i
\end{eqnarray*}
which implies:
\begin{equation}
\lambda = \frac{1}{4 \sqrt{\delta_i}}\sqrt{\sum_{m=1}^M  \frac{d_{S_i,W_k}^{2 \alpha}}{d_{S_i,D_i}^{2 \alpha}} \frac{g^4_{(S_i,D_i),m} }{g^4_{(S_i,W_k),m}} \frac{\sigma^4_{W_k,m}}{\sigma^4_{D_i,m}}}.
\end{equation}
Note that the summation under the square root sign in the above equation can be calculated based on the known parameters of link $i$.  Hence, define:

\begin{equation}
\Gamma_i = \sum_{m=1}^M  \frac{d_{S_i,W_k}^{2 \alpha}}{d_{S_i,D_i}^{2 \alpha}} \frac{g^4_{(S_i,D_i),m} }{g^4_{(S_i,W_k),m}} \frac{\sigma^4_{W_k,m}}{\sigma^4_{D_i,m}}.
\label{Gamma}
\end{equation}

\noindent Then, $\lambda = \frac{1}{4} \sqrt{\frac{{\Gamma_i}} {\delta_i}}$ and 
\begin{equation}
P_{S_i,j} = \sqrt{\frac{{\delta_i}}{\Gamma_i}}\frac{d_{S_i,W_k}^{2 \alpha}}{d_{S_i,D_i}^{\alpha}} \frac{g^2_{(S_i,D_i),j} \sigma^4_{W_k,j}}{g^4_{(S_i,W_k),j}\sigma^2_{D_i,j}}.
\label{Power}
\end{equation}
Putting this into the original optimization problem to find the link capacity yields a capacity for link $i$ of:
\begin{eqnarray*}
    C_i & = & \sqrt{\frac{\delta_i}{\Gamma_i}} \sum_{m=1}^M  \frac{d_{S_i,W_k}^{2 \alpha}}{d_{S_i,D_i}^{2 \alpha}} \frac{g^4_{(S_i,D_i),m} }{g^4_{(S_i,W_k),m}} \frac{\sigma^4_{W_k,m}}{\sigma^4_{D_i,m}} \\
      & = & \sqrt{\delta_i \Gamma_i}
\end{eqnarray*}  
where, as noted earlier, $\Gamma_i$ can be readily computed from the link parameters.  Hence, given the link parameters, the capacity is readily found given the covertness constraint $\delta_i$; since this dependence will be important in the next section where we optimally allocate the covertness constraints across links, we note it explicitly by writing:
\begin{equation}
    C_i(\delta_i) = \sqrt{\delta_i \Gamma_i}.
    \label{C_i}
\end{equation}

\subsection{Finding the Route with Maximum Capacity}
\label{subsec:Het-Opt}
The capacity of a given path $\Pi$ with $H$ hops is the minimum of the capacities over the links in the path:
\begin{equation*}
    C_{\Pi} = \mbox{min}_{i=1,2,\ldots, H} ~C_i(\delta_i).
\end{equation*}
We wish to maximize this capacity subject to an end-to-end covertness constraint.  The optimization problem becomes how to allocate the covertness constraint over the links:
\begin{equation}
    C_{\Pi} = \max_{\delta_1, \delta_2,\ldots, \delta_H}  \mbox{min}_{i=1,2,\ldots, H}~ C_i(\delta_i)
\end{equation}
given the covertness constraint ${\cal D}({\mathbb P}_0 || {\mathbb P}_1) \leq \delta$.  Because of the conditional independence of the observations at the adversary given the presence or absence of a message moving across the network,
\begin{equation*}
    {\cal D}({\mathbb P}_0 || {\mathbb P}_1) = \sum_{i=1}^H {\cal D}({\mathbb P}^{(i)}_0 || {\mathbb P}^{(i)}_1)
\end{equation*}
and thus it is sufficient to employ a link-wise additive covertness constraint; that is,
\begin{equation*}
    \sum_{i=1}^H \delta_i = \delta.
\end{equation*}
Recognizing that the links will all have the same capacity at the optimal point, let $C_{\Pi}$ be that optimal capacity.   Then, using the results of the previous section: for each $i=1,2\ldots, H$, $C_{\Pi}=\sqrt{\delta_i \Gamma_i}$ and thus $\delta_i= \frac{C_{\Pi}^2}{\Gamma_i}$.  Inserting this into the constraint equation yields that the capacity of the path is:
\begin{equation}
    C_{\Pi} = \sqrt{\frac{\delta}{\sum_{i=1}^H \frac{1}{\Gamma_i}}}
\label{Optimal Capacity}
\end{equation}
and hence we want to choose the path that has the smallest $\sum_{i=1}^H \frac{1}{\Gamma_i}$, which is readily done in polynomial time via shortest-path routing with link weight (cost) of $\frac{1}{\Gamma_i}$. Algorithm \ref{alg:cap} (Het-Opt) shows these steps and the incorporation of Dijkstra's algorithm \cite{Dijkstra1959} to find the path and link configurations that optimize the capacity given an end-to-end covertness constraint. Algorithm \ref{alg:cap} (Het-Opt) runs in \(O(N^2)\) for \(N\) friendly nodes, which is the standard complexity for Dijkstra's algorithm.
\RestyleAlgo{ruled}
\begin{algorithm}[t]
\caption{({\bf Het-Opt}) Optimal route and link configuration for maximum capacity with a covertness constraint}\label{alg:cap}
    1. Compute \(\Gamma_i\) for the link between each pair of nodes in the network using (\ref{Gamma}).
    
    2. Find the optimal route that minimizes \(\sum_{i=1}^H \frac{1}{\Gamma_i}\), using Dijkstra's algorithm \cite{Dijkstra1959} while setting the weights of each link as \(\frac{1}{\Gamma_i}\). 

    3. Compute the capacity $C_{\Pi}$ for the optimal path using (\ref{Optimal Capacity}).
    
    4.  For each link $l_i$, $i=1,2,\ldots,H$ in the optimal path $\Pi=(l_1,l_2,\ldots, H)$ returned from Dijkstra's algorithm:
    
    \begin{itemize}
        \item {Calculate the link's (optimal) covertness constraint \\ 
        with $\delta_i= \frac{C_{\Pi}^2}{\Gamma_i}$.}
        \item{Use $\Gamma_i$, $\delta_i$, and the link parameters to find the  \\ optimal power 
        \(P_{S_i,j}\), $j=1,2,\ldots, M$ to assign \\ each mode 
        (and, hence, the degree of mode \\ utilization) for link $i$ 
        using (\ref{Power}).}
        \end{itemize}
    
\end{algorithm}

\subsection{Developing an Algorithm for Comparison:  Maximum Capacity with a Fixed Decision Error Probability (DEP) Per Link}
\label{subsec:comparison}
The algorithm of the previous section and its extensions presented in Section \ref{sec:extensions} are the main contributions of this paper.  However, to establish a viable algorithm for comparison for the numerical results of Section \ref{sec:numerical}, here we adapt the algorithm of \cite{fikadu_tifs_2023} to an end-to-end covertness constraint.

One of the key features of Algorithm 1 (Het-Opt) is the partitioning of the end-to-end covertness constraint $\delta$ over the links to provide the optimal values of $\delta_i, i=1,2,\ldots, H$.  In contrast, in \cite{fikadu_tifs_2023}, the decision error probability (DEP) is constrained to be fixed per link.  We derive an algorithm for comparison to Algorithm 1 (Het-Opt) under this fixed link DEP constraint, as follows.  First, we fix a maximum (very large) number of hops $H$.  Then, for each $h=1,2,\ldots, H$;  (i) set $\delta_i=\delta/h$ as the DEP for the link between each pair of nodes in the network; (ii) with $\delta_i$ given, the capacity of each link can be readily found; (iii) the maximum capacity route with a maximum number of hops $h$ is then readily obtained.  The maximum possible capacity is obtained over $h=1,2,\ldots, H$.  The detailed algorithm is given in Algorithm \ref{alg:cap2} (Per-Link-DEP).


\begin{algorithm}
\caption{({\bf Per-Link-DEP}) Route and link configuration for maximum capacity by extending the fixed decision error probability (DEP) per link approach of \cite{fikadu_tifs_2023} to an end-to-end covertness constraint}\label{alg:cap2}
    1. Compute \(\Gamma_i\) for the link between each pair of nodes in the network using (\ref{Gamma}).

    2. Find the optimal route that maximizes \(\Gamma_i\) for a given maximum hops $h$ using Algorithm \ref{Bellman_Ford_mod}.

    3. Repeat step 2 for different hops \(h \in \{1,2,\ldots,H\}\).

    4. Compute the capacity for different $h$ using (\ref{C_i}).

    5. Pick the route that results in maximum capacity among $h$.
\end{algorithm}

The Algorithm \ref{alg:cap2} (Per-Link-DEP) runs in \(O(N^4)\), for \(N\) friendly nodes, if hops $h$ is considered to be as large as \(N\).
\begin{algorithm}[htp]
\caption{Modified Bellman-Ford Algorithm with Maximum Hops $h$}\label{Bellman_Ford_mod}
1. Initialize the distance for each node to be infinity and set the distance for Source to be zero.\\
2. Perform edge relaxation for each edge in the network. \\
\If {d[v] > d[u] + w(u, v)}{
$d[v]$ ← $d[u] + w(u, v)$

$\pi[v]$ ← $u$
}
Here \(d\) is the distance, \(w(u,v)\) is the weight of the edge from node \(u\) to \(v\), and \(\pi\) is the predecessor of node \(v\).

3. Repeat step 2, $h$ times, storing the distance and predecessors in a matrix for each iteration.

4. Return the final distance and predecessor list. 
\end{algorithm}
\section{Extensions}
\label{sec:extensions}

The optimization in Section \ref{sec:optimization} forms the core result of this paper.  However, two critical extensions for the result to be applied in practice are considered here.

\subsection{Multiple Adversaries}

While ostensibly more complicated, the case of multiple adversaries (Willies) extends naturally from the case of a single Willie with appropriately modified link costs.  The derivation of the link costs to support a polynomial-time routing algorithm to find the optimal route and link configuration parameters in the case of multiple adversaries is included in Appendix A, and numerical results for this important case are included in Section \ref{sec:numerical}.

\subsection{Imperfect Channel State Information}

Suppose that the uncertainty in the knowledge of the channel state information at the adversary can be characterized statistically; for example, it would be reasonable to assume that the Rayleigh fading coefficient to the adversary is unknown by system nodes. Although it will be clear from the below that the approach is readily generalized to any uncertainty distribution, we consider here a channel state estimate with additive Gaussian error.  Hence, let the channel coefficient of mode \(m\), \(g_{(S_i,W_k),m}\), be a complex Gaussian random variable with known mean $v_{(S_i,W_k),m}$ on one of its two components, and variance $\sigma^2$ on each of the components.  Then, \(\sqrt{g_{(S_i,W_k),m}}\) is Rician distributed for a transmitter of link \(i\), \(S_i\) and a Willie \(W_k\) under mode \(m\). The Rician shape factor \(K_{(S_i,W_k),m} =\frac{v_{(S_i,W_k),m}^2}{2\sigma_{(S_i,W),m}^2}\) is then the ratio of the power of the known component $v_{(S_i, W_k),m}$ to that of the scattered components (or estimation error). Letting $K=0$ allows us to model complete uncertainty about the channel fading, whereas letting $K \rightarrow \infty$ retrieve the case considered in Section \ref{sec:optimization} with perfect knowledge of the CSI.



To have the same probabilistic covertness guarantees of Section \ref{sec:model}, we guarantee the covertness constraint on average; that is, the convertness constraint becomes:
\begin{equation*}
\mathbb{E}_{\{g_{(S_i,W_k),m}, m=1,2,\ldots,M\}} \left [ \sum_{m=1}^M \left (\frac{g^2_{(S_i,W_k),m}P_{S_i,m}}{\sigma^2_{W_k,m} d_{S_i,W_k}^{\alpha}} \right)^2 \right ] \leq \delta_i
\end{equation*}
\noindent which implies:
\begin{equation*}
 \sum_{m=1}^M \left (\frac{P_{S_i,m}}{\sigma^2_{W_k,m} d_{S_i,W_k}^{\alpha}} \right)^2  E_{g_{(S_i,W_k),m}}[g^4_{(S_i,W_k),m}] \leq \delta_i.
\end{equation*}
The well-known moments of a Rician random variable yield:
\begin{eqnarray*}
    \mathbb{E}_{g_{(S_i,W_k),m}}\left[g^4_{(S_i,W_k),m}\right] =~~~~~~~~~~~~~~~~~~~~~~~~~~~~~~~~~  \\
    8 \sigma_{(S_i,W),m}^4
    + 8 \sigma_{(S_i,W),m}^2 v_{(S_i,W),m}^2+ v_{(S_i,W),m}^4.
\end{eqnarray*}



\noindent Define $\tau_{(S_i,W),m} = 8 \sigma_{(S_i,W),m}^4
    + 8 \sigma_{(S_i,W),m}^2 v_{(S_i,W),m}^2+ v_{(S_i,W),m}^4$. Then, the covertness constraint employed in Section \ref{sec:optimization} is replaced with:
\begin{equation*}
\sum_{m=1}^M \left (\frac{\tau_{(S_i,W),m}P_{S_i,m}}{\sigma^2_{W_k,m} d_{S_i,W_k}^{\alpha}} \right)^2 \leq \delta_i.
\end{equation*}
Following an analogous derivation to that of Section \ref{sec:optimization}, an optimal polynomial-time routing and link configuration algorithm is obtained with link weights given by:
\begin{equation*}
\Gamma_i = \sum_{m=1}^M  \frac{d_{S_i,W_k}^{2 \alpha}}{d_{S_i,D_i}^{2 \alpha}} \frac{g^4_{(S_i,D_i),m} }{\tau_{(S_i,W),j}^2} \frac{\sigma^4_{W_k,m}}{\sigma^4_{D_i,m}}.
\end{equation*}
\section{Simulation Results and Evaluations}
\label{sec:numerical}
\begin{figure*}[t]
  \begin{subfigure}{0.2\textwidth}
    \includegraphics[width=\linewidth]{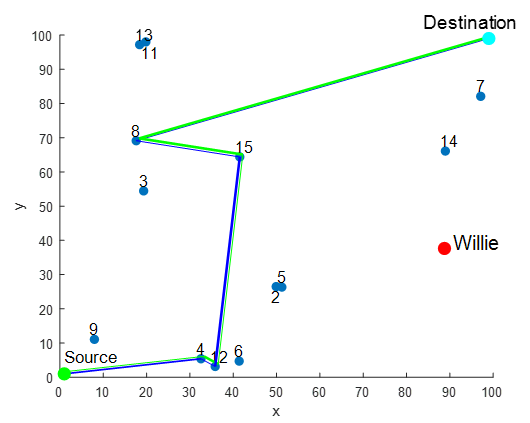}
    \caption{\centering Friendly nodes, N = 15} \label{fig:toy_example_a}
  \end{subfigure}%
  \hspace*{\fill}   
  \begin{subfigure}{0.2\textwidth}
    \includegraphics[width=\linewidth]{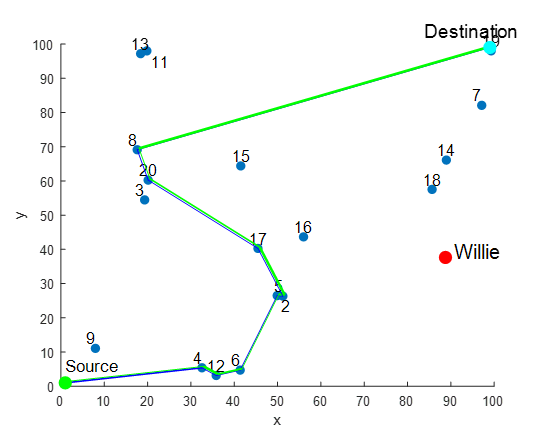}
    \caption{\centering N = 20} \label{fig:toy_example_b}
  \end{subfigure}%
  \begin{subfigure}{0.2\textwidth}
    \includegraphics[width=\linewidth]{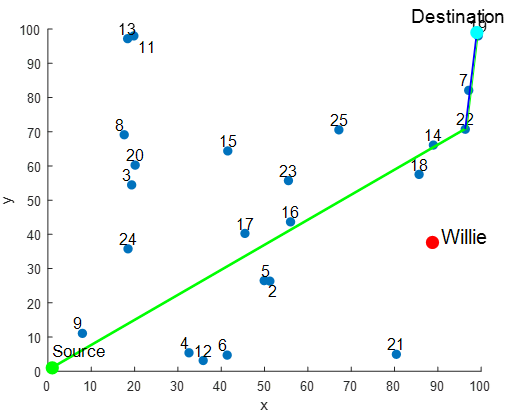}
    \caption{\centering N = 25} \label{fig:toy_example_c}
  \end{subfigure}%
  \begin{subfigure}{0.2\textwidth}
    \includegraphics[width=\linewidth]{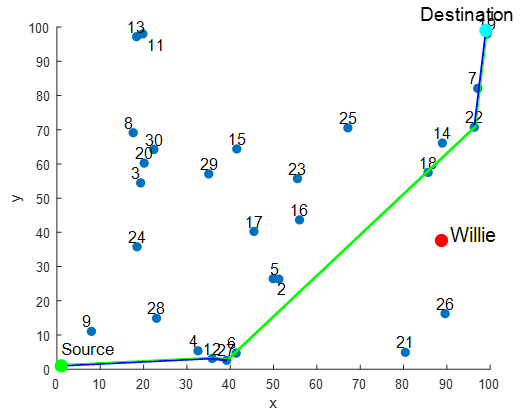}
    \caption{\centering N = 30} \label{fig:toy_example_d}
  \end{subfigure}%
  \begin{subfigure}{0.2\textwidth}
    \includegraphics[width=\linewidth]{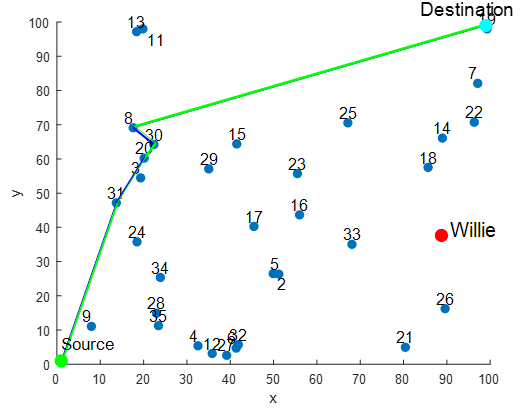}
    \caption{\centering N = 35} \label{fig:toy_example_e}
  \end{subfigure}%

\caption{Sample optimal path using Algorithm 1 (Het-Opt); a given link employs the optimal combination of Mode 1 (AWGN channel, marked with blue), and Mode 2 (fading channel, marked with green).  The thickness of an edge corresponds to the power allocated to that mode for a given link.} \label{fig:toy example}
\end{figure*}
In this section, we provide simulation results to evaluate the performance of the proposed scheme. We first provide single-link simulations to quantify the benefits of using multiple modes in isolation without the link dependencies in a network scenario. Next, results for the multi-hop case are provided to demonstrate: (i) the benefits of employing multiple modes in multi-hop network scenarios; and (ii) the advantage of employing the proposed Het-Opt scheme (see Section 
\ref{subsec:Het-Opt}) versus even well-designed alternatives such as Per-Link-DEP where the covertness constraint is split evenly across the links (see Section \ref{subsec:comparison}).

\subsection{Simulation Set-up}

The simulation setup consists of $N\in \{10,15, \ldots 35\}$ friendly nodes, placed uniformly at random in an area of \(100 \times 100\). Various path loss exponents will be considered below, but the path loss exponent is \(\alpha =2\) unless specified otherwise. The source is placed in one corner of the region at [1,1], and the destination is placed in the opposite corner at [99,99] to maximize the distance that the message must travel in the network and to guarantee that the adversary Willie is located somewhere between the source and destination.  The adversary Willie's location is randomly selected for each iteration. The block length in our simulation is assumed to be $n=500$. We use \(\epsilon=0.01\) as our covertness criterion, where \(\epsilon\) represents the bound on the Kullback-Leibler (KL) divergence per Section \ref{sec:model}. A smaller \(\epsilon\) value indicates a more stringent criterion for covertness, making it harder for an adversary to detect the communication. The covertness criterion per symbol is thus \(\delta = \epsilon/n = 2 \times 10^{-5}\). These covertness settings are obviously very strict and will lead to the small capacities expected at this level of security and adversary capability; for this reason, we plot (\ref{eq:linearized_capacity}) in each case.  Each plot is generated by averaging the results of \(10^4\) randomly generated networks. 

Two modes are considered:  Mode 1 (AWGN channel) and Mode 2 (fading channel). The channel statistics for the AWGN channel are \(\sigma_{w,1}^2 = 1\), and \(\sigma_{D_i,1}^2 \sim \mathcal{U}(1,4)\), where $\sim \mathcal{U}(a,b)$ means the random variable is distributed uniformly between $a$ and $b$; this selection both provides the adversary Willie with an advantage and considers variations in receiver quality at the friendly nodes.   Similarly, for the fading channel, the noise power is \(\sigma_{w,2}^2 = 1\) and \(\sigma_{D_i,2}^2 \sim \mathcal{U}(1,4)\), and a standard normalized Rayleigh channel fading model is employed; that is, \(g_{(S_i,D_i),2} \sim \mathcal{CN}(0, 1)\) and \(g_{(S_i,W),2} \sim \mathcal{CN}(0, 1)\).

\subsection{Single-link Results}

\begin{figure}[htp]
  \begin{subfigure}{0.24\textwidth}
    \includegraphics[width=\linewidth]{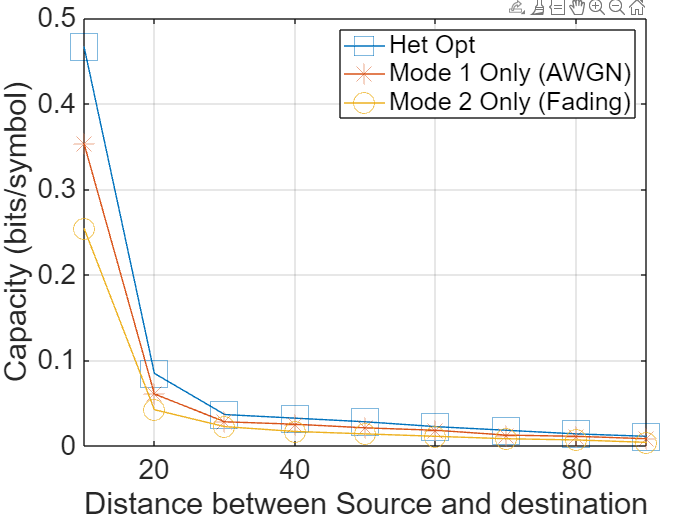}
    \caption{\centering \scriptsize Versus Source to Destination Distance} \label{fig:Single_link1}
  \end{subfigure}%
  \hspace*{\fill}   
  \begin{subfigure}{0.24\textwidth}
    \includegraphics[width=\linewidth]{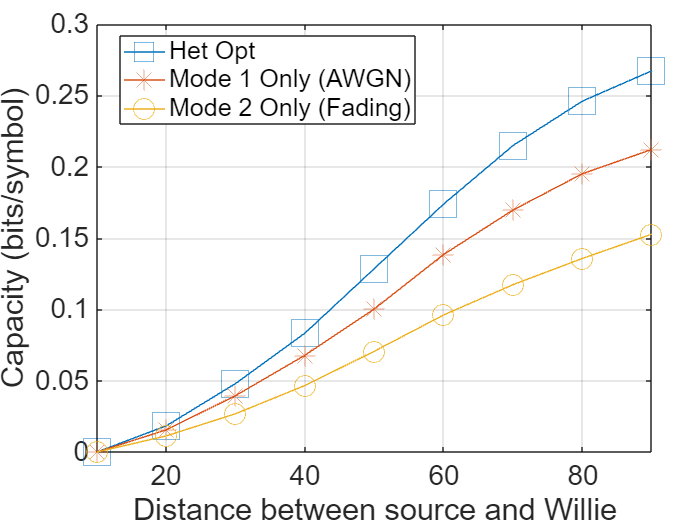}
    \caption{\centering \scriptsize Versus Source to Willie Distance} \label{fig:Single_link2}
  \end{subfigure}%

\caption{Single-link capacity of the proposed optimal combination of modes (Het-Opt) and the capacity when employing either of the modes in isolation.  Note that the gains of Het-Opt are significant here, but will be even further enhanced with the additional freedom (routing) of the multi-hop scenario.} \label{fig:1}
\end{figure}

Figure \ref{fig:1} shows the results for a single-link multi-mode system, where link configuration is performed for different scenarios.
For Figure \ref{fig:Single_link1}, the destination node is placed randomly around the source at a given distance. As the distance between the source and destination increases, the capacity decreases rapidly, regardless of Willie's location. This rapid decline in capacity is due to the attenuation of the covert wireless signals. Clearly, the transmission of covert signals over long distances is hindered without the utilization of multi-hop strategies, as considered in the next section.  Figure \ref{fig:Single_link2} shows the capacity versus distance between the source and Willie. 

In both scenarios shown in Figure 2, the multi-mode scheme exhibits a significant capacity gain compared to the individual utilization of Mode 1 (AWGN channel) or Mode 2 (fading channel).  This highlights the efficacy of our proposed multi-mode approach, which configures a single link to harness the optimal characteristics of both modes.  However, as will be observed in the next section, even larger gains are possible when the Het-Opt algorithm can exploit the degrees of freedom offered by optimal routing in addition to optimal link configuration.

\subsection{Multi-hop Results}

Figure \ref{fig:toy example} illustrates examples of paths chosen by Algorithm 1 (Het-Opt) for different numbers of friendly nodes. For example, Figure \ref{fig:toy_example_a} shows the path from the source to the destination in the presence of 15 nodes, with the network capacity being 0.0778. Blue represents Mode 1 (AWGN channel), while green represents Mode 2 (fading channel). The thickness of the edges indicates the proportion of power allocated to each mode on a given link. For example, in Figure \ref{fig:toy_example_a}, the edge between node 8 and the destination is mostly green, indicating that more power is allocated to Mode 2 (fading channel). In contrast, the other edges are predominantly blue, indicating a higher use of Mode 1 (AWGN channel).  In general, edges relatively close to Willie (the adversary) employed by Het-Opt tend to dedicate more power to Mode 2 (fading channel), while edges further away from Willie allocate more power to Mode 1 (AWGN channel).  This can be explained, as follows:  Mode 1 (AWGN channel) links whose sources are close to Willie are known to be poor and thus are avoided, whereas Mode 2 (fading channel) links close to Willie may be in a good or bad state, and Het-Opt will employ those links only if they are in a good state.

This pattern becomes more prominent in Figure \ref{fig:toy_example_c} and Figure \ref{fig:toy_example_d}, where the link between the source and node 22 and the link between node 12 and node 22 use predominantly Mode 2 (fading channel).  As the optimal path in the presence of more friendly nodes changes, the capacity increases to 0.1174, 0.1994,  0.2107, and 0.3422 for Figure \ref{fig:toy_example_b} through Figure \ref{fig:toy_example_e}.

\begin{figure}[htp]
  \begin{subfigure}{0.23\textwidth}
    \includegraphics[width=\linewidth]{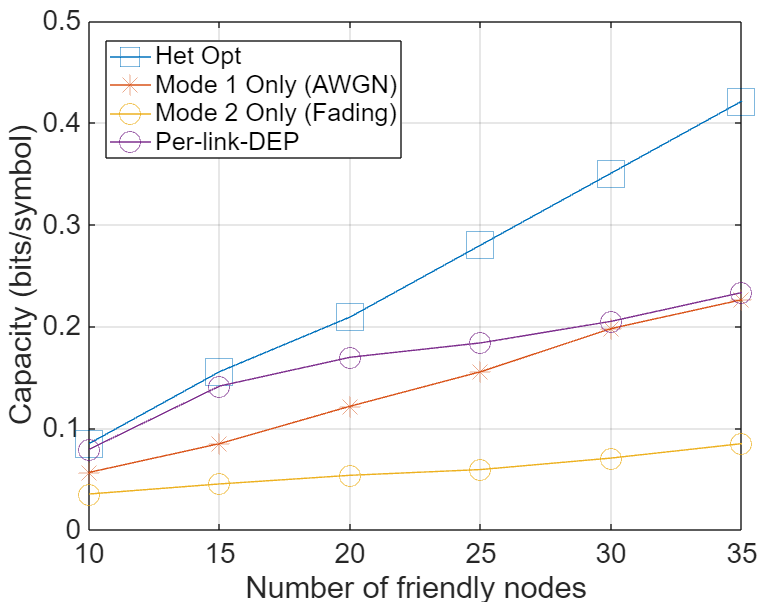}
    \caption{\centering path-loss, alpha = 2} \label{fig:alpha=2}
  \end{subfigure}%
  \hspace*{\fill}   
  \begin{subfigure}{0.24\textwidth}
    \includegraphics[width=\linewidth]{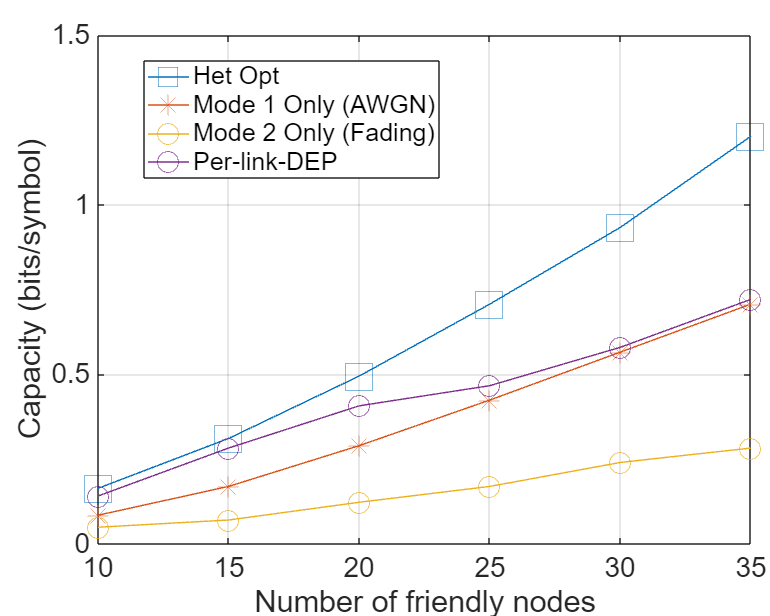}
    \caption{\centering path-loss, alpha = 4} \label{fig:alpha=4}
  \end{subfigure}%

\caption{Capacity for the proposed scheme (Het-Opt), Mode 1 (AWGN channel) only, Mode 2 (fading channel) only, and Per-link-DEP with the maximum number of hops set to 10. } \label{fig:2}
\end{figure}

Figure \ref{fig:2} shows capacity versus the number of friendly nodes for the proposed Het-Opt scheme, Mode 1 (AWGN channel) only, Mode 2 (fading channel) only, and the Per-link-DEP algorithm (see Algorithm 2 in Section \ref{subsec:comparison}) developed for comparison.  
As the number of friendly nodes increases, the capacity for a given source to the destination increases regardless of Willie's location.  The Het-Opt scheme outperforms the schemes that employ optimal routing and power assignment for a single mode, for varying path loss, as shown in Figure \ref{fig:alpha=2} and Figure \ref{fig:alpha=4}.  As expected, the gains in the multi-hop scenario, where Het-Opt can exploit more degrees of freedom while performing optimal routing and multimode link configuration, are more significant than those in the single-hop case studied in the previous section.  The performance of the Per-link-DEP method is close to Het-Opt for small numbers of friendly nodes but, as the number of nodes increases, its performance decreases compared to Het-Opt and approaches that of Mode 1 (AWGN channel). This indicates that although the Per-link-DEP method is effective, it lacks adaptability when many options are available in dynamic wireless networks.

\begin{figure}[htp]
\centering
\includegraphics[scale=0.26]{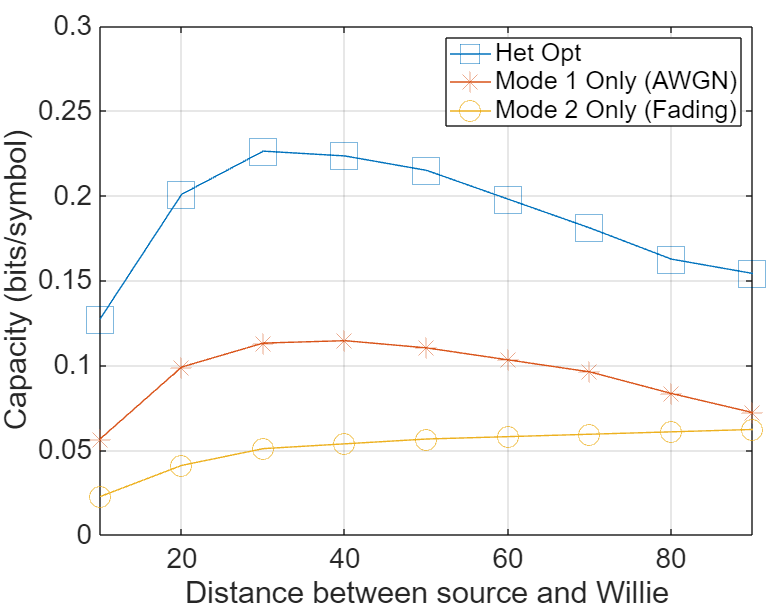}
\caption{Capacity versus distance between source and adversary for the proposed scheme (Het-Opt), Mode 1 (AWGN channel) only, and Mode 2 (fading channel) only.}
\label{fig:Multilink1}
\end{figure}

Figure \ref{fig:Multilink1} shows the capacity versus distance between the source and the adversary for various algorithms. Het-Opt outperforms the other algorithms significantly again, as expected.  Whereas the capacity of Het-Opt initially increases as the adversary Willie moves away from the source, it eventually starts to decrease as the adversary nears the destination, hence forcing Het-Opt and Mode 1 (AWGN channel) only to use links that are near the adversary.  While the performance of the Mode 2 (fading channel) only algorithm improves with increasing distance, it still greatly under-performs Het-Opt in all cases.  


\begin{figure}[htp]
\centering
\includegraphics[scale=0.3]{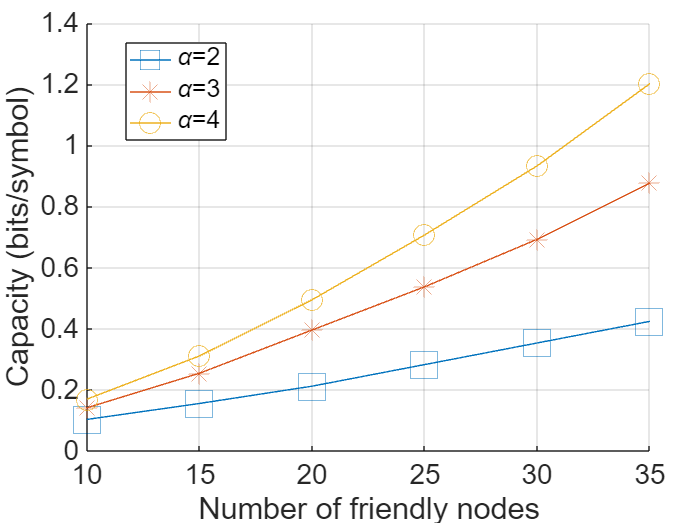}
\caption{Comparison of the capacity of Het-Opt for different values of the path loss exponent $\alpha$.}
\label{fig:path-loss}
\end{figure}

Figure \ref{fig:path-loss} shows the Het-Opt capacity for different values of path-loss, \(\alpha=2,3,4\).  In most wireless scenarios, increasing the path loss would decrease the capacity, but here, in the case of covert communication, higher path loss also makes it more challenging for adversaries to detect the covert signal - especially when the routing algorithm is able to choose favorable routes that stay away from the adversary Willie (e.g., a listening tower or adversary unit from which we would like to avoid detection).  This can allow the source to use higher power and hence a more complex modulation scheme to increase capacity without being detected.

\begin{figure}[t]
  \begin{subfigure}{0.25\textwidth}
    \includegraphics[width=\linewidth]{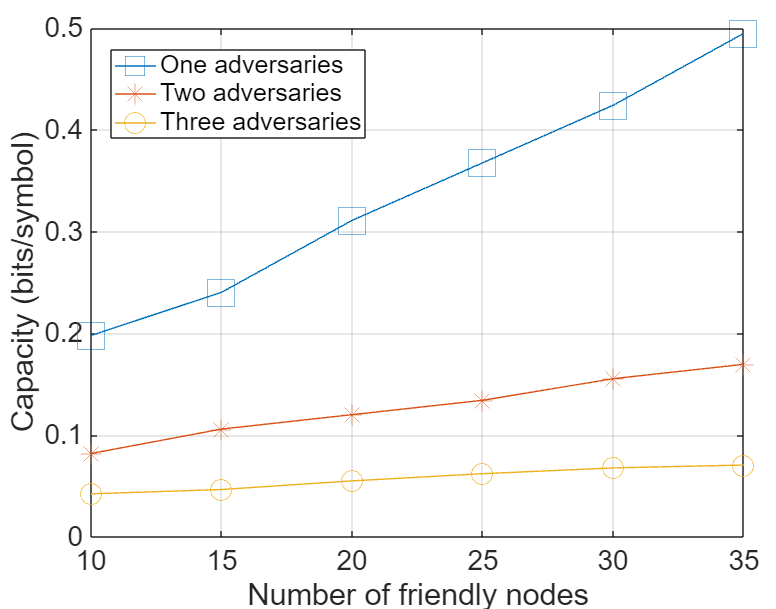}
    \caption{\centering Performance of Algorithm 1 (Het Net) with randomly placed adversaries} \label{fig:Multilink4}
  \end{subfigure}%
  \hspace*{\fill}   
  \begin{subfigure}{0.25\textwidth}
    \includegraphics[width=\linewidth]{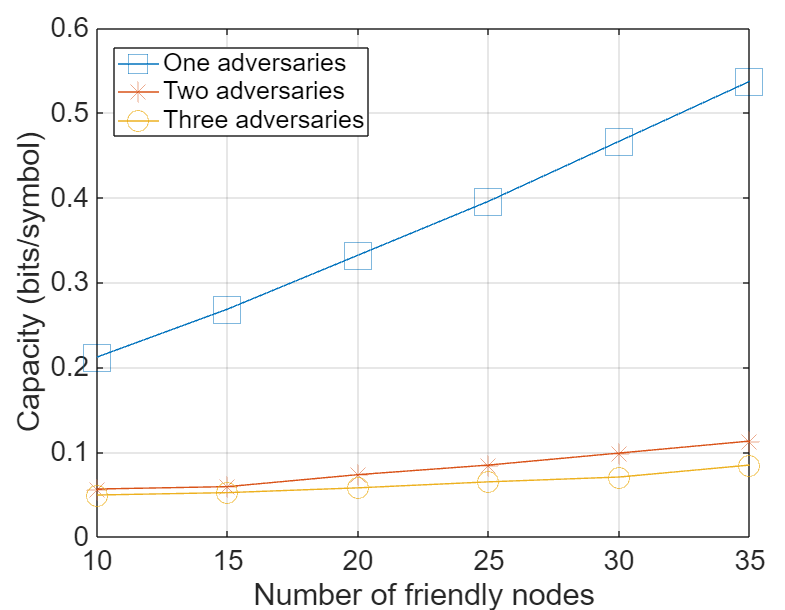}
    \caption{\centering Performance of Algorithm 1 (Het Net) with intelligently placed adversaries} \label{fig:Multilink3}
  \end{subfigure}%
\vspace{1cm} 
\centering
\begin{subfigure}{0.25\textwidth}
    \includegraphics[width=\linewidth]{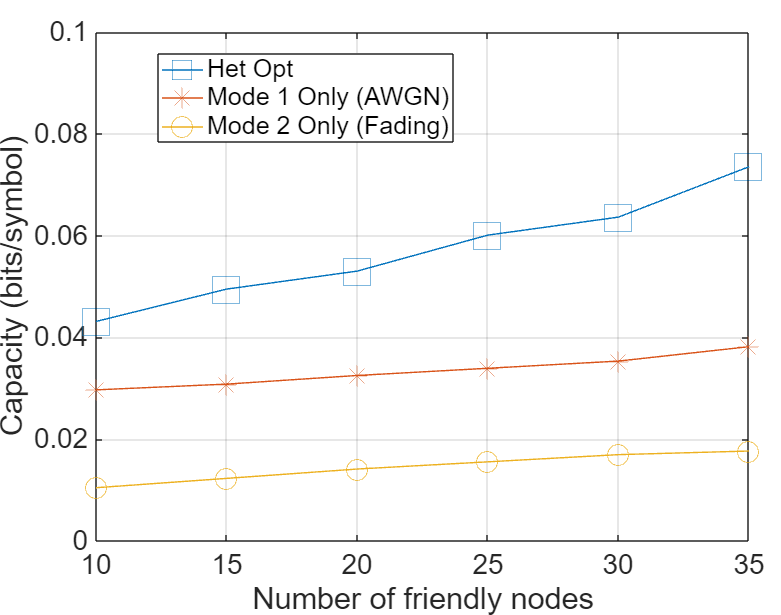}
    \caption{\centering Comparison between different algorithms in the presence of three adversaries} \label{fig:comparison for three willies}
  \end{subfigure}%
\caption{Capacity versus the number of friendly nodes for multiple adversaries} \label{fig:3}
\end{figure}

Figure \ref{fig:3} shows the numerical results evaluating the performance of Algorithm 1 (Het-Opt) in the presence of multiple adversaries.  In Figure \ref{fig:Multilink3}, adversaries are placed randomly.  However, in Figure \ref{fig:Multilink3}, the adversaries are placed ``intelligently'', which means that the adversaries know each other's locations, so a minimum distance constraint is applied in their separation.  Not surprisingly, as the number of adversaries increases, the capacity for the end-to-end optimal route decreases even in the presence of a high number of friendly nodes, and a significant drop in capacity is observed when the adversaries are placed intelligently compared to the case of their random placement. Figure \ref{fig:comparison for three willies} shows the capacity in the presence of three randomly placed adversaries and demonstrates that the proposed Het-Opt scheme significantly outperforms single-mode approaches even in the presence of more than one adversary. It can also be seen from the figure that, as the number of adversaries increases, the curve becomes flatter reflecting a reduced effect of more friendly nodes in the presence of more adversaries.

\section{Conclusion}
\label{sec:conclusion}

We have considered optimal routing in wireless networks to maximize the capacity under an end-to-end covertness constraint when nodes potentially have multiple modalities available for transmission.  An efficient polynomial-time routing algorithm has been derived that provides both the optimal multi-hop route from a source to the destination and the optimal configuration of nodes along that route.  Numerical results demonstrate that the proposed Het-Opt method, where multiple modes are used simultaneously in an optimal fashion, outperforms the use of single modes and a method motivated by previous work that allocates the end-to-end covertness constraint equally across links to readily allow for efficient routing.  Moreover, the proposed (Het-Opt) scheme is effective in the presence of multiple adversaries in finding the optimal capacity route.

\vspace*{0.2in}

\begin{center} {APPENDIX A:  OPTIMIZATION IN THE PRESENCE OF MULTIPLE ADVERSARIES}
\end{center}

In this appendix, we extend the optimization to the case of multiple adversaries (Willies).  
Consider first the optimization on a single link.  The problem becomes: 
\begin{equation*}
\mbox{max}_{P_{S_i,1}, P_{S_i,2},\ldots, P_{S_i,M}} \sum_{m=1}^M  \frac{g^2_{(S_i,D_i),m} P_{S_i,m}}{2\sigma^2_{D_i,m} d_{S_i,D_i}^{\alpha}} 
\end{equation*}
such that
\begin{equation*}
\sum_{m=1}^M \left (\sum_{k=1}^{N_w} \frac{g^2_{(S_i,W_k),m}P_{S_i,m}}{\sigma^2_{W_k,m} d_{S_i,W_k}^{\alpha}} \right)^2 \leq \delta_i
\end{equation*}
Define the Lagrangian:
\begin{eqnarray*}
    {\cal L}((P_{S_i,1}, P_{S_i,2},\ldots, P_{S_i,M}),\lambda) = ~~~~~~~~~~~~~~~~~~~~~~~~~~~~~~~~~~ \\
      \sum_{m=1}^M  \frac{g^2_{(S_i,D_i),m} P_{S_i,m}}{2\sigma^2_{D_i,m} d_{S_i,D_i}^{\alpha}} + \lambda \left (\delta_i - \sum_{m=1}^M \left (\sum_{k=1}^{N_w}\frac{g^2_{(S_i,W_k),m} P_{S_i,m}}{\sigma^2_{W_k,m} d_{S_i,W_k}^{\alpha}} \right)^2 \right )  \\
\end{eqnarray*}
Taking the partial derivatives with respect to $P_{S_i,1}, P_{S_i,2}, \ldots P_{S_i,M}$ and setting equal to zero yields the necessary conditions ($j=1,2,\ldots,M$):
\begin{equation*}
P_{S_i,j} = \frac{g^2_{(S_i,D_i),j}}{4 \lambda \sigma^2_{D_i,j} d_{S_i,D_i}^{\alpha} \left (\sum_{k=1}^{N_w}\frac{g^2_{(S_i,W_k),j} }{\sigma^2_{W_k,j} d_{S_i,W_k}^{\alpha}} \right)^2 }
\end{equation*}
Substituting into the constraint equation results in:
\begin{equation*}
\sum_{m=1}^M  \frac{ \left (\sum_{k=1}^{N_w} \frac{g^2_{(S_i,W_k),m}}{\sigma^2_{W_k,m} d_{S_i,W_k}^{\alpha}} \right)^2 g^4_{(S_i,D_i),m}}{16 \lambda^2 \sigma^4_{D_i,m} d_{S_i,D_i}^{2 \alpha} \left (\sum_{k=1}^{N_w}\frac{g^2_{(S_i,W_k),m} }{\sigma^2_{W_k,m} d_{S_i,W_k}^{\alpha}} \right)^4 } = \delta_i
\end{equation*}
Solving for $\lambda$ yields:
\begin{equation*}
    \lambda = \frac{1}{4 \sqrt{\delta_i}} \sqrt{\sum_{m=1}^M \frac{g^4_{(S_i,D_i),m}}{\sigma^4_{D_i,m} d_{S_i,D_i}^{2 \alpha} \left (\sum_{k=1}^{N_w}\frac{g^2_{(S_i,W_k),m} }{\sigma^2_{W_k,m} d_{S_i,W_k}^{\alpha}} \right)^2}}
\end{equation*}
As in the case of a single Willie, everything inside the large square root is readily available from the link parameters; hence, define
\begin{equation*}
\Gamma^{(N_w)}_i = \sum_{m=1}^M \frac{g^4_{(S_i,D_i),m}}{\sigma^4_{D_i,m} d_{S_i,D_i}^{2 \alpha} \left (\sum_{k=1}^{N_w}\frac{g^2_{(S_i,W_k),m} }{\sigma^2_{W_k,m} d_{S_i,W_k}^{\alpha}} \right)^2}
\end{equation*}
Then, $\lambda = \frac{1}{4} \sqrt{\frac{\Gamma^{(N_w)}_i}{\delta_i}}$ and
\begin{equation*}
P_{S_i,j} = \sqrt{\frac{\delta_i}{\Gamma^{(N_w)}_i}}{\frac{g^2_{(S_i,D_i),j}}{\sigma^2_{D_i,j} d_{S_i,D_i}^{\alpha} \left (\sum_{k=1}^{N_w}\frac{g^2_{(S_i,W_k),j} }{\sigma^2_{W_k,j} d_{S_i,W_k}^{\alpha}} \right)^2 }}~~~~~~~~~~~
\end{equation*}
The rest of the routing optimization follows identically to the case of a single Willie, with $\Gamma_i$ replaced with $\Gamma_i^{(N_w)}$ throughout.


\end{document}